\documentclass{elsart}
\usepackage[dvips]{graphics,graphicx,color}
\usepackage{indentfirst,latexsym,amsfonts,amssymb,amsmath}

\begin{document}

\begin{frontmatter}

\title{Microscopic modelling of defects production and their annealing 
after irradiation in silicon for HEP particle detectors}
\author[iftm]{S. Lazanu\thanksref{cor}},
\author[univ]{I. Lazanu} and
\author[fl]{M. Bruzzi}

\address[iftm]{National Institute for Materials Physics, POBox MG-7, Bucharest-Magurele, 
Romania}
\address[univ]{University of Bucharest, Department of Nuclear Physics, POBox MG-11, 
Bucharest-Magurele, Romania}
\address[fl]{Universita di Firenze, Dipartamento di Energetica, Via S. Marta 3, 50139- 
Florence, Italy}
\thanks[cor]{corresponding author: fax: +40-21-4930267, e-mail:lazanu@alpha1.infim.ro}
\begin{abstract}
In this contribution, the production of defects in radiation fields and their evolution 
toward equilibrium in silicon for detector uses has been modelled. In the quantitative 
model developed, the generation rate of primary defects is calculated starting from the 
projectile - silicon interaction and from  recoil energy redistribution in the lattice. 
Vacancy-interstitial annihilation, interstitial migration to sinks, divacancy and 
vacancy-impurity complex ($VP$, $VO$, $V_2O$, $C_iO_i$ and $C_iC_s$) formation are 
considered.

The results of the model support the experimental available data. 

The correlation between the initial material parameters, temperature, irradiation and 
annealing history is established. The model predictions could be a useful clue in 
obtaining harder materials for detectors at the new generation of accelerators or for 
space missions.

\medskip
61.70.At: Defects formation and annealing processes.\\
29.40.Pe: Semiconductor detectors.
\end{abstract}

\end{frontmatter}

\section{Introduction}

Silicon is the most used semiconductor material for detectors in particle physics 
experiments. The principal obstacles to long term operation in the extreme radiation 
conditions are the changes in detector parameters, consequence of the modifications in 
material properties during and after irradiation, ultimately related to the defects 
induced this way in the lattice.

A point defect in a crystal is an entity that causes an interruption in the lattice 
periodicity. In this paper, the terminology and definitions in agreement with M. Lannoo 
and J. Bourgoin \cite{1} are used in relation to defects.

We denote the displacement defects, vacancies and interstitials, as primary point defects, 
prior to any further rearrangement.

The microscopic model developed in this contribution is able to explain quantitatively, 
without free parameters, the production of primary defects in silicon, and their evolution 
toward equilibrium because in silicon, these defects are essentially unstable and interact 
via migration, recombination, and annihilation or produce other defects.

The generation rate of primary defects is considered in a large range of values, and is 
calculated starting from the projectile - silicon nucleus interaction, and from the 
analytical extensions of the Lindhard theory \cite{2}.

Silicon used for detectors in high energy physics is n-type high resistivity ($1\div 6 
K\Omega\cdot cm$) phosphorus doped FZ material. The concentrations of interstitial oxygen  
$O_i$ and substitutional carbon $C_S$  in silicon are strongly dependent on the growth 
technique. In high purity Float Zone Si, oxygen interstitial concentrations are around 
$10^{15}$ cm$^{-3}$, while in Czochralski Si these concentrations can reach values as high 
as $10^{18}$ cm$^{-3}$. Because Czochralski silicon is not available in detector grade 
quality, an oxygenation technique developed at BNL produces Diffusion Oxygenated Float 
Zone in silicon, obtaining a $O_i$ concentration of the order $5\cdot10^{17}$ cm$^{-3}$. 
These materials can be enriched in substitutional carbon up to [C$_i] \approx 1.8 \cdot 
10^{16}$ cm$^{-3}$.

Vacancy-interstitial annihilation, interstitial migration to sinks, divacancy and 
vacancy-impurity complex ($VP$, $VO$, $V_2O$, $C_iO_i$ and $C_iC_s$) formation are 
considered.

Model predictions are compared with experimental data.

The correlation between the initial material parameters, temperature, irradiation and 
annealing history is established. 

\section{Production of primary defects and their kinetics}

The basic assumption of the present model is that the primary defects, vacancies and 
interstitials, are produced in equal quantities and are uniformly distributed in the 
material bulk. They are produced by the incoming particle, or as a consequence of the 
subsequent collisions of the primary recoil in the lattice.

Prior to the irradiation process, in silicon there are thermally generated defects (only 
Frenkel pairs are considered).

After the irradiation, the following stable defects have been identified in silicon (see 
References \cite{1,3}): $Si_i$, $VP$, $VO$, $V_2$, $V_2O$, $C_iO_i$, $C_i$ and $C_iC_s$.

The pre-existing thermal defects and those produced by irradiation, as well as the 
impurities, are assumed to be randomly distributed in the solid. An important part of the 
vacancies and interstitials annihilate. The sample contains certain concentrations of 
impurities, which can trap interstitials and vacancies respectively, and form stable 
defects.

Vacancy-interstitial annihilation, interstitial migration to sinks, divacancy, vacancy and 
interstitial impurity complex formation are considered. The role of phosphorus, oxygen and 
carbon is taken into account, and the following stable defects:  $VP$, $VO$, $V_2$, 
$V_2O$, $C_iO_i$, $C_i$ and $C_iC_s$ are considered. Other possible defects as $V_3O$, 
$V_2O_2$, $V_3O_3$ \cite{4}, are not included in the present model.

The following picture describes in terms of chemical reactions the mechanisms of 
production and evolution of the defects considered in the present paper:
\begin{equation}                                                        
V+I\ _{\overleftarrow{G}} ^{\underrightarrow{K_1}}\text{annihilation}
\end{equation}
\begin{equation}                                                       
I\stackrel{K_2}{\rightarrow } \text{sinks}
\end{equation}
\begin{equation}							
V+O\ _{\overleftarrow{K_4}} ^{\underrightarrow{K_3}}\ VO
\end{equation}
$VO$ is the $A$ centre.
\begin{equation}							
V+P\ _{\overleftarrow{K_5}} ^{\underrightarrow{K_3}}\ VP
\end{equation}
$VP$ is the $E$ centre.
\begin{equation}						
V+V\ _{\overleftarrow{K_6}} ^{\underrightarrow{K_3}}\ V_2
\end{equation}
\begin{equation}						
V+VO\ _{\overleftarrow{K_{7}}} ^{\underrightarrow{K_{3}}}\ V_2O
\end{equation}
\begin{equation}
I+C_s\stackrel{K_1}{\rightarrow }  C_i
\end{equation}
\begin{equation}
C_i+O_i\stackrel{K_8}{\rightarrow } C_iO_i
\end{equation}
\begin{equation}
A+I\stackrel{K_9}{\rightarrow }  O
\end{equation}
\begin{equation}
C_i+C_s\stackrel{K_8}{\rightarrow }  C_iC_s
\end{equation}

The reaction constants $K_i$ (i = 1, $3 \div 8$) have the general form:
\begin{equation}
K_i=C\nu \exp \left( -E_i/k_BT\right)
\end{equation}
with $\nu$ the vibration frequency of the lattice, $E_i$ the associated activation energy 
and $C$ a numerical constant that accounts for the symmetry of the defect in the lattice.

The reaction constant related to the migration of interstitials to sinks could be 
expressed as:
\begin{equation}
K_2=\alpha \nu \lambda ^2\exp \left( -E_2/k_BT\right)                                                 
\end{equation}
with $\alpha$ the sink concentration and $\lambda$ the jump distance.

The system of coupled differential equations corresponding to the reaction scheme 
(1)$\div$ (10) cannot be solved analytically and a numerical procedure was used.

The following values of the parameters have been used: $E_1$=$E_2$ = 0.4 eV, $E_3$ = 0.8 
eV, $E_4$ = 1.4 eV, $E_5$ = 1.1 eV, $E_6$ = 1.3 eV, $E_7$ = 1.6 eV, $E_8$ = 1.7 eV, $\nu$ 
= 10$^{13}$ Hz, $\lambda$ = 10$^{15}$ cm$^2$, $\alpha$ = 10$^{10}$ cm$^{-2}$.

The generation term  is the sum of two components:
\begin{equation}
G=G_R+G_T
\end{equation}
where $G_R$ accounts for the generation by irradiation, and $G_T$ for thermal generation.

The concentration of the primary radiation induced defects per unit fluence ($CPD$) in 
silicon has been calculated using the explicit formula (see details, e.g. in references 
\cite{5,6}):
\begin{equation}
CPD	\left(E\right)= \frac{N_{Si}}{2E_{Si}} \int \sum _{i} \left( \frac{d\sigma}{d\Omega} 
\right)_{i,Si} L(E_{Ri})_{Si} d\Omega=\frac{1}{N_A} \frac{N_{Si}A_{Si}}{2E_{Si}} 
NIEL\left(E\right)
\end{equation}
where $E$ is the kinetic energy of the incident particle, $N_{Si}$ is the atomic density 
in silicon, $A_{Si}$ is the silicon atomic number, $E_{Si}$ - the average threshold energy 
for displacements in the semiconductor, $E_{Ri}$  - the recoil energy of the residual 
nucleus produced in interaction $i$, $L(E_{Ri})$ - the Lindhard factor that describes the 
partition of the recoil energy between ionisation and displacements and 
$(d\sigma/d\Omega)_i$ - the differential cross section of the interaction between the 
incident particle and the nucleus of the lattice for the process or mechanism $i$, 
responsible in defect production. $N_A$ is Avogadro's number. The formula gives also the 
relation with the non ionising energy loss ($NIEL$). 

Due to the important weight of annealing processes, as well as to their very short time 
scale, $CPD$ is not a measurable physical quantity.

If the anisotropy of the silicon lattice is not considered, the simplifying hypothesis of 
random distribution of $CPD$ for all particles could be introduced: the identity of the 
particle is lost after the primary interaction, and two different particles could produce 
the same generation rate ($G_R$) of vacancy-interstitial pairs if the following condition:
\begin{equation}
G_R=[\left(CPD\right)_{part.a}\left(E_1)\right]\cdot\Phi_{part.a}(E_1)=[\left(CPD\right)_{
pa
rt.b}\left(E_2)\right]\cdot\Phi_{part.2(E_2)}
\end{equation}
is fulfilled.
Here, $\Phi$ is the flux of particles ($a$) and ($b$) respectively, and $E_1$ and $E_2$ 
their corresponding kinetic energies.

\section{Results and discussion}
\subsection{Comparison between model predictions and experimental data}

The model predictions have been compared with experimental measurements. A difficulty in 
this comparison is the insufficient information in published papers regarding the 
characterisation of silicon, and the irradiation parameters and conditions for most of the 
data.

It was underlined in the literature \cite{7} that the ratio of $VO$  to $VP$  centres in 
electron irradiated silicon is proportional to the ratio between the concentrations of 
oxygen and phosphorus in the sample. For electron irradiation, in Reference \cite{8} a 
linear dependence of the $V_2$  versus $VO$  centre concentration has been found 
experimentally. In the present paper, the ratios of concentrations of  $V_2$ to $VO$  
centres and $VO$  to $VP$  ones has been calculated respectively in the frame of the 
model, for the material with the characteristics specified in Reference \cite{8}, and 
irradiated with 12 MeV electrons, in the conditions of the above mentioned article. The 
time dependence of these two ratios is represented in Figures 1a and 1b; annealing is 
considered both during and after irradiation. It could be seen that for the ratio of $V_2$  
and $VO$  concentrations, the curves corresponding to different irradiation fluences are 
parallel, while the ratio of $VO$  to $VP$  concentrations is fluence independent, in the 
interval $2\cdot 10^{13}\div 5.5\cdot 10^{14}$ cm$^{-2}$, in good agreement with the 
experimental evidence. The ratio between  $V_2$ and $VO$  concentrations is determined by 
the generation of primary defects by irradiation, while the ratio between $VO$  and  $VP$  
concentrations is determined by the concentrations of oxygen and phosphorus in silicon. 

\begin{figure}[ht]
\centering
\includegraphics[width=.6\textwidth]{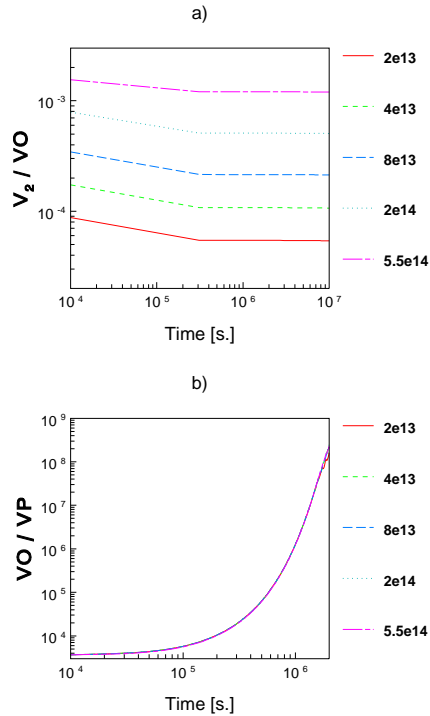}
\caption{\small{Time dependence for a)$V_2/VO$ and b)$VO/VP$ concentrations calculated for 
silicon with $1.4\cdot 10^{14}$ P/cm$^{3}$, $5\cdot 10^{17}$ O/cm$^{3}$, $3\cdot 10^{15}$ 
C/cm$^{3}$, after 12 MeV electron irradiation, with the flux $5.8\cdot 10^{10}$ 
e/cm$^{2}\cdot s$, up to the fluences: $2\cdot 10^{13}$ e/cm$^{2}$, $4\cdot 10^{13}$ 
e/cm$^{2}$, $8\cdot 10^{13}$ e/cm$^{2}$, $2\cdot 10^{14}$ e/cm$^{2}$,  $2\cdot 10^{13}$ 
e/cm$^{2}$, $5.5\cdot 10^{14}$ e/cm$^{2}$, followed by relaxation (see reference 
\cite{8}.}}
\label{f1}
\end{figure}

Our estimations are also in agreement with the measurements presented in reference 
\cite{9}, after electron irradiation, where defect concentrations are presented as a 
function of the time after irradiation. In Figure 2, both measured and calculated 
dependencies of the $VO$  and $VP$  concentrations are given. The irradiation was 
performed with 2.5 MeV  electrons, up to a fluence of  $3\cdot 10^{16}$ cm$^{-2}$. The 
dependencies put in evidence the important role played by the carbon-related defects. The 
relative values are imposed by the arbitrary units of experimental data.

\begin{figure}[ht]
\centering
\includegraphics[width=.55\textwidth]{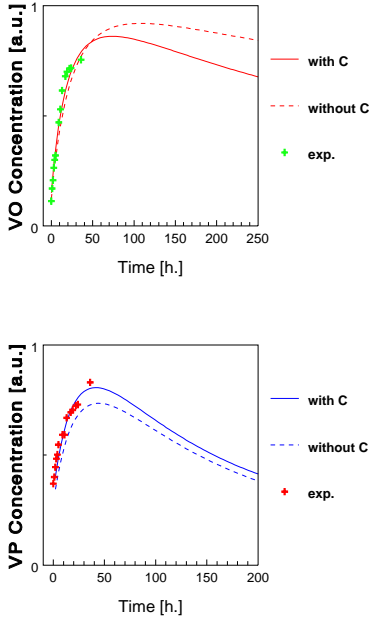}
\caption{\small{Time dependence of $VO$ and $VP$ concentrations after electron 
irradiation: crosses - experimental data from reference \cite{9}; continuous line present 
model calculations; dashed line - without the consideration of carbon contribution to 
defect formation.}}
\label{f2}
\end{figure}

A good agreement has also been obtained between the absolute values of concentrations of 
$VP + V_2$  and $C_iC_s$  centres predicted by the model, and the experimental results 
after 1 MeV  neutron irradiation at a total fluence of $5.67\cdot 10^{13}$ cm$^{-2}$, 
reported in reference \cite{10}. The calculated $1.5\cdot 10^{13}$ cm$^{-3}$  and 
$4.1\cdot 10^{12}$ cm$^{-3}$ concentrations for $VP + V_2$   and  $C_iC_s$  respectively, 
are in accord with the values of  $1.1\cdot 10^{13}$ cm$^{-3}$ and  $3.8\cdot 10^{12}$ 
cm$^{-3}$ , measured experimentally. For the  $VO$ concentration, a poorer concordance has 
been obtained.

The formation and time evolution of stable defects depends on various factors, e.g. the 
concentrations of impurities pre-existent in the sample, the rate of generation, and the 
temperature and the irradiation history.

\subsection{The variation of the concentration of stable defects in respect to the 
irradiation rate and  the influence of oxygen}

The effect of oxygen in irradiated silicon has been a subject of intensive studies in 
remote past. In the last decade a lot of studies have been performed to investigate the 
influence of different impurities, especially oxygen and carbon, as possible ways to 
enhance the radiation hardness of silicon for detectors in the future generation of 
experiments in high energy physics - see, e.g. references \cite{11,12}. Some authors 
consider that these impurities added to the silicon bulk modify the formation of 
electrically active defects, thus controlling the macroscopic device parameters. 
Empirically, it is considered that if the silicon is enriched in oxygen, the capture of 
radiation-generated vacancies is favoured by the production of the pseudo-acceptor complex 
vacancy-oxygen. Interstitial oxygen acts as a sink for vacancies, thus reducing the 
probability of formation of the divacancy related complexes, associated with deeper levels 
inside the gap.

Two types of silicon have been considered in the model calculations: "standard" material, 
with the following doping concentrations: $10^{14}$ cm$^{-3}$  phosphorous content,  
$2\cdot 10^{15}$ cm$^{-3}$  oxygen, and  $3\cdot 10^{15}$ cm$^{-3}$  carbon, and 
"oxygenated" material, with the same concentrations of phosphorous and carbon, but with 
200 higher oxygen content,  $4\cdot 10^{17}$ cm$^{-3}$ .

These two types of materials are considered to be exposed to continuous irradiation, with 
$G_R$  varying from $5\cdot 10^9 VI$pairs/s, to 50 $VI$pairs/s, from half to half order of 
magnitude, at 20$^o$C, during 10 years. 

Thermal generation has been considered each time in the calculations. The results are 
illustrated in Figure 3. 
\begin{figure}[ht]
\centering
\includegraphics[width=.8\textwidth]{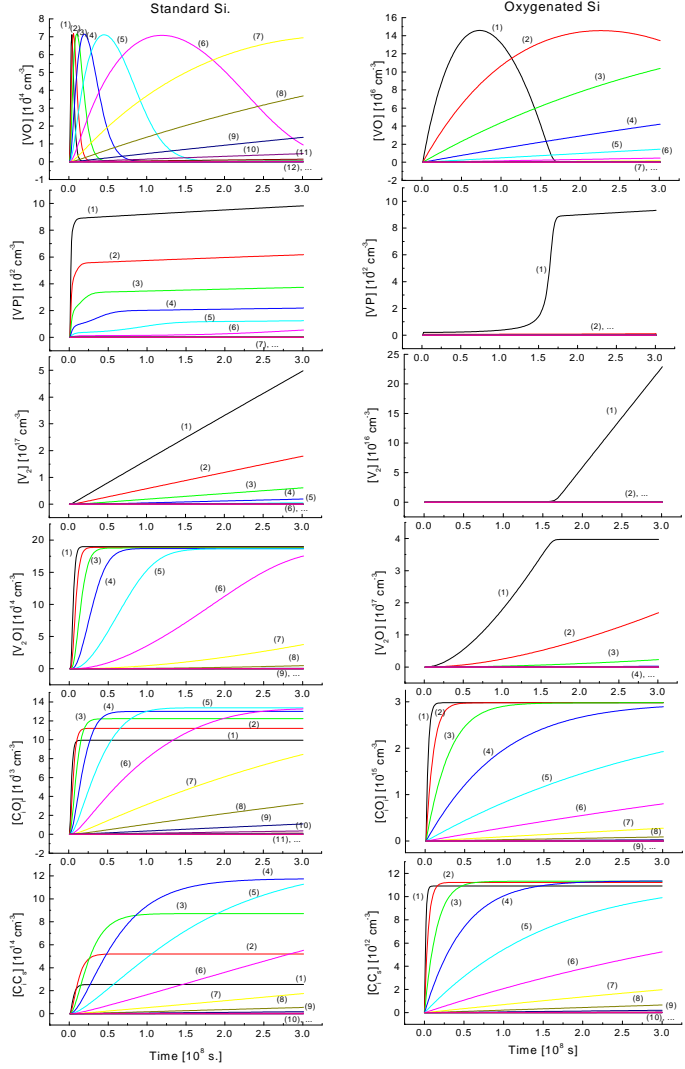}
\caption{\small{Influence of irradiation rate on the time evolution of defect 
concentrations. The values of $G_R$ are decreasing from $5\cdot 10^9$ VI pairs/s - curve 
(1) to 50 VI pairs/s - curve (17), from half to half order of magnitude.}}
\label{f3}
\end{figure}
The highest generation rate corresponds here to a value 10 times higher than that 
estimated for the forward position in the tracker silicon detector at the new LHC hadron 
accelerator (determined from integration over the energy range of interest of the 
convoluted function between simulated hadron flux spectra \cite{13} and the $CPD$), while 
the smallest generation rate corresponds to a value 10 times smaller than the rate 
produced by protons from cosmic rays, in an orbit near the Earth, at about 380 Km, as 
measured by the AMS Collaboration \cite{14}.

It could be seen that the presence of oxygen slows down the formation of all defects 
related to oxygen and vacancy, i.e. $VO$, $VP$, $V_2$, $V_2O$ and $C_iO_i$, and increases 
the rate of formation of $C_iC_s$  centres. It could also be observed that the highest 
concentrations attained by the  $VO$ centres during this period, as well as those 
corresponding to the $V_2O$  ones increase, from "standard" to "oxygenated" silicon, in 
the same ratio as the corresponding oxygen content. Because the process of formation of 
both divacancies and $V_2O$  centres is, in an important measure, slowed down in 
oxygenated silicon, and the maximum values of divacancy concentrations are reduced with 
the increase of oxygen concentration, the generation current in the depleted regions of 
p-n junction detectors could be decreased by oxygen addition, especially at smaller rates 
than those associated with curves (1).

\subsection{The effect of the temperature}

Both "standard" and "oxygenated" silicon are considered to be exposed to similar 
continuous irradiations, characterised by a generation rate $G_R = 5\cdot 10^7 VI$pairs/s, 
during ten years, at temperatures between -20$^o$C and 20$^o$C. Thermal generation has 
been considered for each temperature. The results are represented in Figure 4 for the 
concentrations of $VO$, $V_2$ and $V_2O$ centres respectively. 
\begin{figure}[ht]
\centering
\includegraphics[width=.8\textwidth]{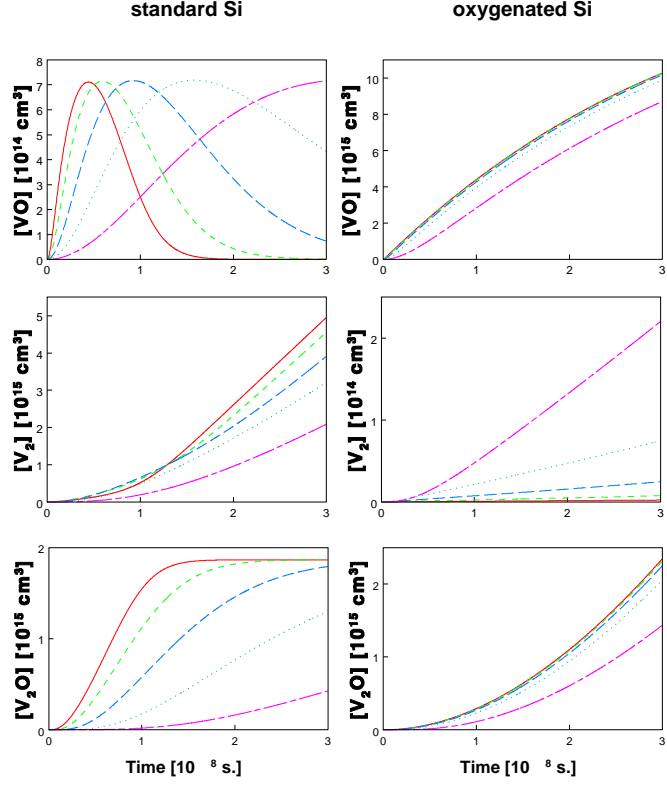}
\caption{\small{Time dependence of the concentrations $VO$, $V_2$ and $V_2O$ centres, in 
the conditions of continuous irradiation with G$_R$ = $5\cdot 10^7$ VI pairs/s, in 
"standard" and "oxygenated" silicon, for 20$^o$C (red), 10$^o$C (green), 0$^o$C (blue), - 
10$^o$C (cyan) and -20$^o$C (magenta) respectively.}}
\label{f4}
\end{figure}
While in "standard" silicon the decrease of the temperature slows down the rate of $VO$ 
formation, a much weaker dependence has been found in silicon enriched in oxygen. 

For the $V_2O$ centres, the maximum concentrations correspond to the highest temperature, 
20$^o$C, and similar values are found at the end of the time interval considered, although 
the introduction rate of this defect has different behaviours in the two types of silicon 
considered.  

In contrast to oxygen related centres, the $V_2$ concentration is more sensitive to 
temperature in "oxygenated" silicon, where its values are much lower in respect to the 
corresponding ones for "standard" material. More, the concentration of $V_2$ centres 
increases with the decrease of the temperature from 20 to -20$^o$C in "oxygenated" 
silicon, and decreases in the "standard" material.

\subsection{The correlation between the history of the irradiation process and defect 
concentrations}

In some previous papers \cite{15,16}, we demonstrated in concrete cases the importance of 
the sequence of irradiation processes, considering that the same total fluence can be 
attained in different situations: the ideal case of instantaneous irradiation, irradiation 
in a single pulse followed by relaxation, and respectively continuos irradiation process. 
As expected, after instantaneous irradiation the concentrations of defects are higher in 
respect with "gradual" irradiation.

After this analysis, the specific importance of the irradiation and annealing history 
(initial material parameters, type of irradiation particles, energetic source spectra, 
flux, irradiation temperature, measurement temperature, temperature and time between 
irradiation and measurement) on defect evolution must be underlined.

Now, we present comparatively in Figure 5 the time evolution of stable defects in standard 
silicon, irradiated during $10^7$ s, when $5\cdot 10^7 VI$ pairs/s   have been generated 
(continuous line), and during $2\cdot 10^7$ s, with $2.5\cdot 10^7 VI$ pairs/s    (dashed 
line), respectively. The values have been chosen to produce the same amount of vacancy - 
interstitial pairs. The behaviour of stable defect concentrations is analysed after 
irradiation. It could be seen that a shorter and stronger irradiation conduces to higher 
concentrations of $V_2$ and $VP$ centres in the asymptotic limit, with important 
consequences for the leakage current, due to the position in the band gap of silicon of 
these defect levels.
\begin{figure}[ht]
\centering
\includegraphics[width=.7\textwidth]{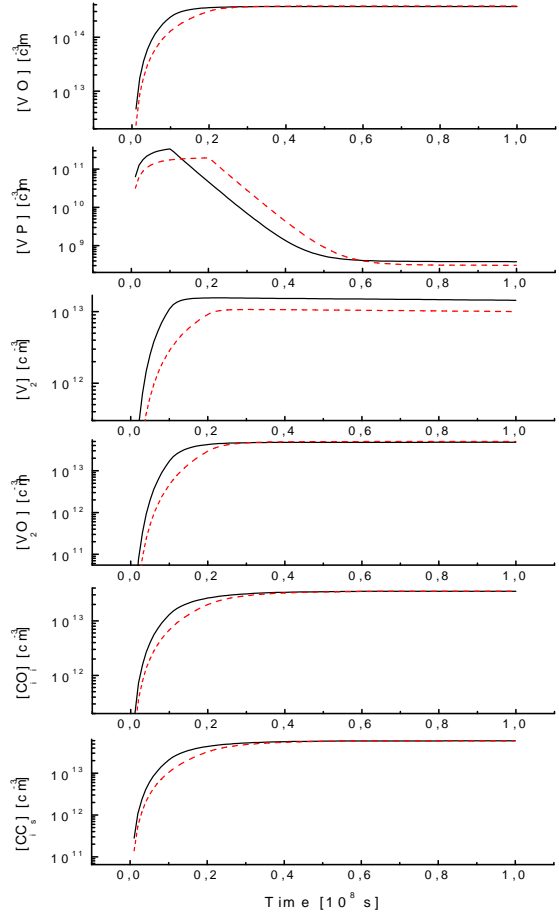}
\caption{\small{Time evolution of the concentrations of (up to down): $VO$, $VP$, $V_2$, 
$V_2O$, $C_iO_i$, and $C_iC_s$ in "standard" Si, at 20$^o$C for two types of irradiation:  
$10^7$ s. irradiation with $G_R = 5\cdot 10^7$  VI pairs/s (continuous line), and $2\cdot 
10^7$ s. $G_R$ = $2.5\cdot 10^7$ VI/s (dashed line) respectively.}}
\label{f5}
\end{figure}

Another scenario that could be of interest is the comparison of the effects of continuous 
and pulsed irradiation, with the same integral rate of primary defect production. The time 
evolution of the concentrations of stable defects is represented in Figure 6. The effects 
of continuous irradiation of standard silicon, at 20$^o$C, with a rate of $2.5\cdot 10^7 
VI$ pairs/s  (continuous line) are compared with pulsed one: $10^7$ s.  irradiation, with 
$5\cdot 10^7 VI$ pairs/s, and $10^7$ s. relaxation. As in the previous case (Figure 5), 
the concentration of $VP$ centres follows the best the irradiation type, while for the 
other defects the succession irradiation - relaxation is almost not observable.
\begin{figure}[ht]
\centering
\includegraphics[width=.7\textwidth]{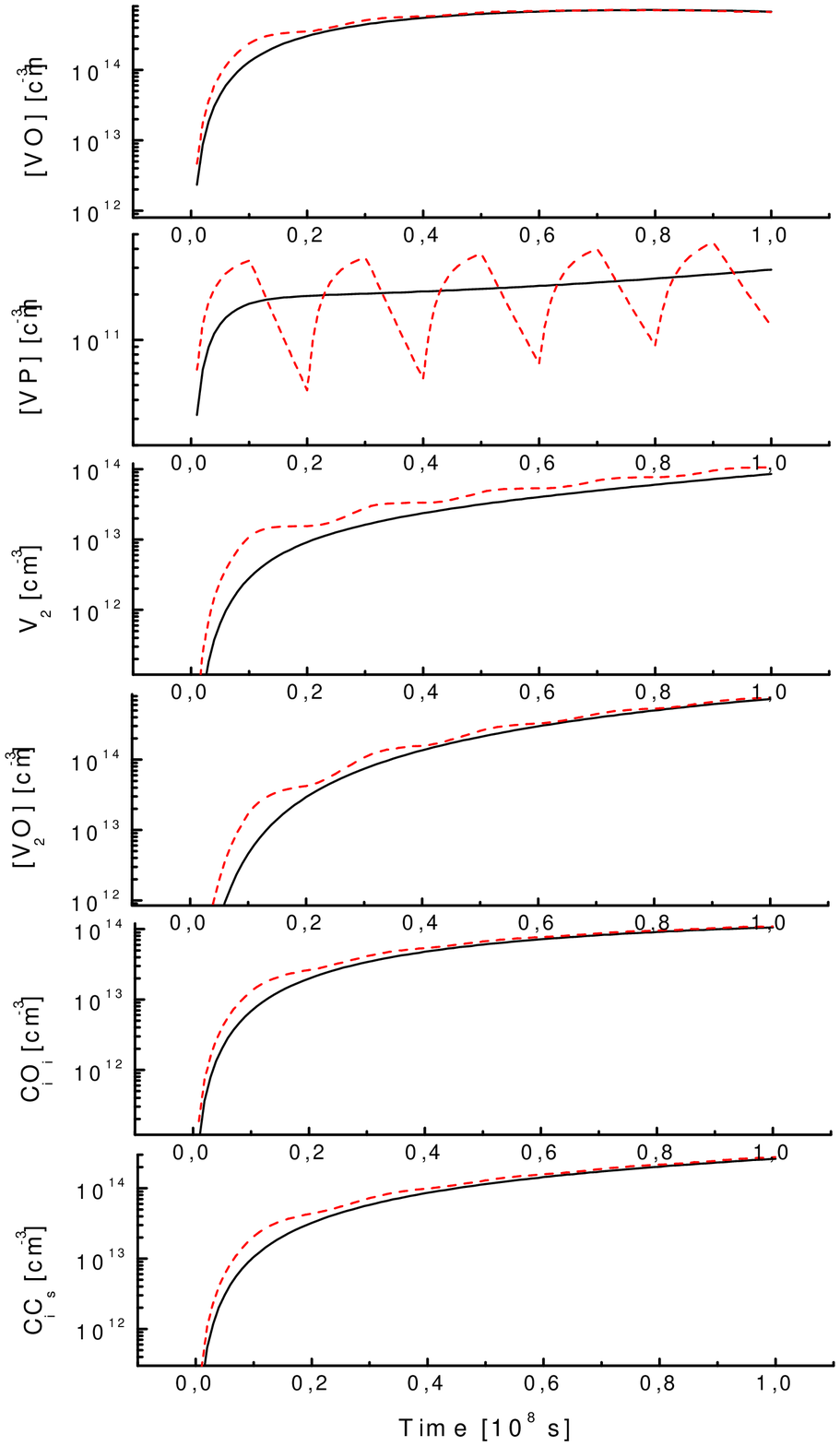}
\caption{\small{Time evolution of the concentrations of (up to down): $VO$, $VP$, $V_2$, 
$V_2O$, $C_iO_i$, and $C_iC_s$  for two types of irradiation: continuous line - continuous 
irradiation with  $2.5\cdot 10^7$ VI pairs/s during 10$^8$ sec, (continuous line), and 
rectangular pulses $10^7$ s. irradiation with $G_R$ = $5\cdot 10^7$ VI/s, and $10^7$ s 
relaxation.}}
\label{f6}
\end{figure}

\section{Summary and conclusions}
The production of defects in radiation fields, in silicon for detector uses, and their 
evolution toward equilibrium, has been microscopically modelled in the frame of a 
quantitative model, without free parameters. Vacancy-interstitial annihilation, 
interstitial migration to sinks, divacancy and vacancy-impurity complexes ($VP$; $VO$; 
$V_2O$, $C_iO_i$ and $C_iC_s$) formation is considered.

The model supports the experimental data and confirms the different studies performed to 
investigate the influence of oxygen in the enhancement of the radiation hardness of 
silicon for detectors. The $VO$  defects in oxygen enriched silicon are favoured in 
respect to the other stable defects, so, for detector applications it is expected that the 
leakage current decreases after irradiation. At high oxygen concentrations, this defect 
saturates starting from low fluences at high generation rates of defects.

The generation rate of primary defects, calculated starting from the projectile - silicon 
nucleus interaction and from the extended Lindhard theory, is considered in a large range 
of values. 

The obtained results suggest the importance of the conditions of irradiation, temperature 
and annealing history on defect kinetics.

\end{document}